\documentclass[12pt]{article}
\usepackage{graphicx}

\newcommand\pubnumber{}
\newcommand\pubdate{October 1, 2015}

\def\grenoble{LPSC, Universit\'e Grenoble-Alpes, CNRS/IN2P3, Grenoble, France}
\def\ILL{Institut Laue Langevin, 53, Rue Horowitz, 38000 Grenoble, France}

\def\Title#1{\begin{center} {\Large #1 } \end{center}}
\def\Author#1{\begin{center}{ \sc #1} \end{center}}
\def\Address#1{\begin{center}{ \it #1} \end{center}}

\newcommand\pubblock{\rightline{\begin{tabular}{l} \pubnumber\\
         \pubdate  \end{tabular}}}
\newenvironment{Abstract}{\begin{quotation}  }{\end{quotation}}
\newenvironment{Presented}{\begin{quotation} \begin{center} 
             PRESENTED AT\end{center}\bigskip 
      \begin{center}\begin{large}}{\end{large}\end{center} \end{quotation}}

\newcommand{\uddu}{$\uparrow\downarrow\downarrow\uparrow$}

\def\beq{\begin{equation}}
\def\eeq#1{\label{#1}\end{equation}}
\def\eeqn{\end{equation}}

\def\beqa{\begin{eqnarray}}
\def\eeqa#1{\label{#1}\end{eqnarray}}
\def\eeqan{\end{eqnarray}}

\newcommand{\eqref}[1]{(\ref{#1})}

\let\bar=\overbar

\def\etal{{\it et al.}}

\def\Dslash{\not{\hbox{\kern-4pt $D$}}}
\def\dslash{\not{\hbox{\kern-2pt $\del$}}}

\def\msb{{\bar{\ssstyle M \kern -1pt S}}}

\begin{document}
\begin{titlepage}
\pubblock

\vfill
\Title{Constraining short-range spin-dependent forces with polarized helium 3 at the Laue-Langevin Institute}
\vfill
\Author{Mathieu Guigue}
\Address{\grenoble}
\Author{David Jullien}
\Address{\ILL}
\Author{Alexander K. Petukhov}
\Address{\ILL}
\Author{Guillaume Pignol}
\Address{\grenoble}
\vfill
\begin{Abstract}
We have searched for a short-range spin-dependent interaction mediated by a hypothetical light scalar boson with CP-violating couplings to the neutron using the spin relaxation of hyperpolarized $^3$He. 
The walls of the $^3$He cell would generate a depolarizing pseudomagnetic field.
%We did not see any anomalous spin relaxation and we report the limit for interaction ranges $\lambda$ between 1~$\mu$m and 100~$\mu$m: $g_sg_p \lambda ^2 \leq 4.9\times 10^{-28}\, ( 95~\%\,  \mathrm{C.L.})$, where $g_s$($g_p$) are the (pseudo)scalar coupling constant, improving the previous best limit by one order of magnitude.
\end{Abstract}
\vfill
\begin{Presented}
Twelfth Conference on the Intersections of Particle and Nuclear Physics (CIPANP2015),\\
Vail Marriott Mountain Resort, Vail, Colorado, USA. 
\end{Presented}
\vfill
\end{titlepage}
\def\thefootnote{\fnsymbol{footnote}}
\setcounter{footnote}{0}
%
%%%%%%%%%%%%%%%%%%%%%%%%%%%%%%%%%
\section{Theoretical motivations}

Theories beyond the Standard Model (SM) of particle physics generally predict the existence of new phenomena at energies well below the electroweak scale.
These new undiscovered particles with masses below 1~eV are called WISPs, which stands for Weakly Interacting Slim Particles.

Since these light particles can arise from the spontaneous breaking of some global symmetry (in the case of spin-0 particles), they are associated with very high energy scales.
A well-motivated example is the hypothetical QCD Axion which arises from the global $U(1)$ Peccei-Quinn symmetry, introduced to solve to the strong-CP problem \cite{Kim2010}.
Even if the mass of this particle is supposed very light (sub-eV), the probe energy scale are very large since the symmetry breaking scale parameter $f_a$ lies between $10^9~\mathrm{GeV}$ and $10^{12}~\mathrm{GeV}$.
Unlike the Axion, axion-like particles have no relation between their masses, the symmetry breaking scale and their coupling to the SM particles.
Other than spin-0 ALPs, new light spin-1 bosons can arise from broken hidden $U(1)$ symmetries and they do no decouple from SM particles even in the limit of vanishing mass.% \cite{Fayet1990}.

%Since they are serious Dark Matter candidates, they are actively searched  for \cite{Ringwald2012,Arias2012}.
%Using the fact that the WISPs behaves like a coherent oscillating field, the \textit{haloscope} experiments, such as ADMX \cite{Asztalos2010}, aim to convert these oscillations into microwave photons using the WISP's coupling to photons.
%Astrophysical objects such as the Sun or supernovae could thermally produce a large number of these particles.
%This radiation would induce a new channel of energy loss which will change the time evolution of these objects.
%Stringent constraints have been set with this astrophysical argument.
%\textit{Helioscope} experiments try to detect on Earth the flux of WISPs emitted from these objects by converting them into X ray photons.

%Since all the experiments described above rely on a astrophysical source, it is interesting to have laboratory experiments which are independent of such sources.
%The coupling between WISP-photon coupling can be probed with \textit{light-shining-through-a-wall} experiments.
%Regarding the coupling to fermions, one way to investigate is the search for \textit{fifth forces} \cite{Antoniadis2011}.
One way to investigate the coupling of these bosons to fermions is the search for \textit{fifth forces} \cite{Antoniadis2011}.
When the interaction between two fermions mediated by the exchange of a light boson  does not depend on the spin of the two fermions, the force derives from a Yukawa potential with a typical interaction range $\lambda = \frac{\hbar c}{m_0 c^2}$.
For a $0.1~\rm{eV}$ boson, the corresponding range of the interaction is $2~\rm{\mu m}$.
Since it corresponds to macroscopic distances, numerous experiments have been realized to probe this fifth force.

Now, instead of  pure scalar couplings which give a Yukawa potential, one can consider a pseudoscalar coupling betwen the boson and the fermions.
The interaction becomes \textit{spin-dependent} and cannot be probe by fifth force experiments.
We considered a interaction potential between a spin and a (unpolarized) mass \cite{Dobrescu2006}
\begin{equation}\label{eq:dipmonointeraction}
 V=g_s g_p  \frac{\hbar \widehat{\sigma}\cdot\widehat{r}}{8\pi Mc}\left(\frac{m_{\phi}}{r}+\frac{1}{r^2}\right)\exp (-m_{\phi} r) \equiv -\vec{\mu}\cdot \vec{b_a} 
\end{equation}
where $\hbar\widehat{\sigma}/2$ is the spin of the probe particle, $M$ the mass of the polarized particle, $g_s$ and $ g_p$ the coupling constant at the vertices of polarized and unpolarized particles corresponding to a scalar and a pseudoscalar interactions. 
The potential \eqref{eq:dipmonointeraction} can be interpreted as a pseudomagnetic field generated by the unpolarized particle on the polarized particle.

Let us consider a $^3$He polarized gas contained in a glass cell.
The source of the interaction acting on the $^3$He is the nucleons in the walls of the cell.
Then each helium spin will locally probe an effective pseudomagnetic field 
\begin{equation}
b(x) = g_{s}g_{p} \frac{\lambda\hbar}{2m\gamma}e^{-x/\lambda}\left( 1-e^{-d/\lambda}\right),
\label{eq:champ-pseudo-total}
\end{equation}
with $\gamma /2\pi = 32.4~\rm{Hz/\mu T}$  the gyromagnetic ratio of the $^3$He atoms and $d$ the thickness of the cell walls.%\cite{Flowers1993}
The motion of the spins in this inhomogeneous pseudomagnetic field will induce a anomalous depolarization channel, in addition to the usual ones.

In Section II, we will present the principle of measurement.
Then the setup will be described in Section III.
In Section IV, the analysis procedure will be presented 
Finally, the constraints obtained up to now will be shown in Section V.

%%%%%%%%%%%%%%%%%%%%%%%%%%%%%%%%%%%%%%
\section{Principle of the measurement}

We consider the case of a gas of $^3$He polarized atoms with a gyromagnetic ratio $\gamma$ contained in a glass cell of volume $V$ and immersed in a holding magnetic field $\vec{B}_0 = B_0 \vec{e}_z$.
The Larmor precession frequency of the spins is $\omega _0 = \gamma B_0$.
In the case of polarized $^3$He gas with a high polarization $P$ ($P(t=0) \approx 70~\%$), the gas depolarizes as
\begin{equation}
P(t) = P(t=0)e^{-\Gamma _1 t},
\label{eq:evolution-P}
\end{equation}
with $\Gamma _1$ the longitudinal relaxation rate.
Three main mechanisms are expected to contribute to the relaxation: collisions between $^3$He atoms, collisions with the cell walls and particles motion in an inhomogeneous magnetic field. 
The first contribution $\Gamma _{dd}$ depends on the frequency of atoms collisions and is proportional to the pressure of the gas \cite{Newbury1993}.
The second contribution $\Gamma _w$ does not depend on the pressure and the polarization of the gas nor on the holding field value.
The last contribution $\Gamma _m$ corresponds to a perturbation of the spins by their motion in an inhomogeneous magnetic field.

In the case of an arbitrary geometry cell immersed in a magnetic field with an arbitrary spatial profile, the relaxation rate can be estimated \cite{Guigue2014} using the Redfield theory with the relation
\begin{equation}
\Gamma _m = D\frac{\langle (\vec{\nabla} b_{\perp})^2\rangle}{B_0^2},
\label{eq:adiabdiff}
\end{equation}
where $D$ is the diffusion coefficient ($D\approx 1.84~\rm{cm^2/s}$ in normal condition ) and $\langle (\vec{\nabla} b_{\perp})^2\rangle$ corresponds to the average over the cell volume of the squared transverse gradients. %\cite{Barbe1974}
This relation is valid in the case of gas with a pressure close to $1~\rm{bar}$, immersed into several $\rm{\mu T}$ holding field which is our case. 
Generally, the magnetic inhomogeneities (small compared with $B_0$) can be written as a function of the polarization $P$ and the pressure $p$ as
\begin{equation}
\vec{b} = \vec{b}_{\mathrm{ext}} + \vec{b}_{\mathrm{coil}} (B_0) + \vec{b}_{\mathrm{cell}} (P,p) ,
\label{eq:realmagfield}
\end{equation} 
with $P$ the polarization of the gas and $p$ its pressure.
The term $\vec{b}_{\mathrm{ext}}$ corresponds to the magnetic inhomogeneities induced by the apparatus environment and $\vec{b}_{\mathrm{coil}} (B_0)$ to the one created by the holding field generator and so is proportional to $B_0$.
The last contribution $\vec{b}_{\mathrm{cell}} (P,p)$ is due to the fact that at high pressure and polarization, the $^3$He gas behaves as a magnet and generates an inhomogeneous magnetic field proportional to $p$ and $P$.

Therefore for a given pressure, using \eqref{eq:adiabdiff} and \eqref{eq:realmagfield}, the total longitudinal relaxation rate $\Gamma _1$ can be written as
\begin{equation}
\Gamma _1 =  a + \frac{b}{B_0} + \frac{c}{B_0^2} + \frac{dP}{B_0} + \frac{eP}{B_0^2} + \frac{fP^2}{B_0^2} ,
\label{eq:Gamma1-explicite}
\end{equation}
with $a$, $b$, $c$, $d$, $e$ and $f$ real coefficients.
The relaxation contributions induced by walls and atoms collisions are contained in the coefficient $a$.

In the case of a scalar-pseudoscalar interaction between a $^3$He polarized gas and the nucleons in the cell walls, the relaxation rate induced by the motion of the spins into a pseudomagnetic field \eqref{eq:champ-pseudo-total} can be derived using the Redfield theory \cite{Guigue2015}.
Let us consider spherical cells with a radius $R$.
The relaxation rate induced by the pseudomagnetic field generated by the walls of such a cell can be written as
\begin{equation}
\Gamma _{NF} = \left( \gamma b_a\right) ^2 \frac{\lambda ^3}{DR}\frac{1}{(1+\phi _{\lambda}^2)^2} \times \left( \sqrt{\frac{2}{\phi _{\lambda}}}\left( 1-\phi _{\lambda}\left( \phi_{\lambda}-2\right)\right) +\phi_{\lambda}^2-3\right) ,
\label{eq:Gamma1NF-formulefinale}
\end{equation}
with $b_a = g_{s}g_{p} \frac{\lambda\hbar}{2m\gamma}\left( 1-e^{-d/\lambda}\right)$ and $\phi _{\lambda} = \gamma B_0 \lambda ^2/D$.
Moreover, for $\phi _{\lambda} \ll 1$, Eq. \eqref{eq:Gamma1NF-formulefinale} simplifies into
\begin{equation}
\Gamma _{NF} = \left( \gamma b_a\right) ^2 \frac{\lambda}{R}\sqrt{\frac{2\lambda ^2}{D\omega _0}}.
\label{eq:GammaNF_simple}
\end{equation}
In this regime, the behavior of the relaxation rate with respect to the holding field is very different from the classical relaxation contributions \eqref{eq:Gamma1-explicite}.
The existence of a new short-range spin-dependent interaction can then be extracted from measurements of the relaxation rate at different values of the holding magnetic field and polarization.

A first test experiment \cite{Petukhov2010} measuring the spin longitudinal depolarization rate as a function of the applied field $B_0$ was performed in 2010 to demonstrate the sensitivity of the method. 
A new dedicated experiment is set up at the Institute Laue-Langevin (ILL), improving both (i) the magnetic environment of the experiment and (ii) the measurement of the decay of polarization.

%%%%%%%%%%%%%%%%%%%%%%%%%%%
\section{Experimental setup description}

Our apparatus is designed to provide a large range of possible holding magnetic field $B_0$ between $3~\rm{\mu T}$ and $300~\rm{\mu T}$ in which the $^3$He gas can depolarized.
The polarization can be measured using dedicated instruments.

\subsection{Magnetic setup}

\begin{figure}[htb]
	\centering
		\includegraphics[width=0.470\textwidth]{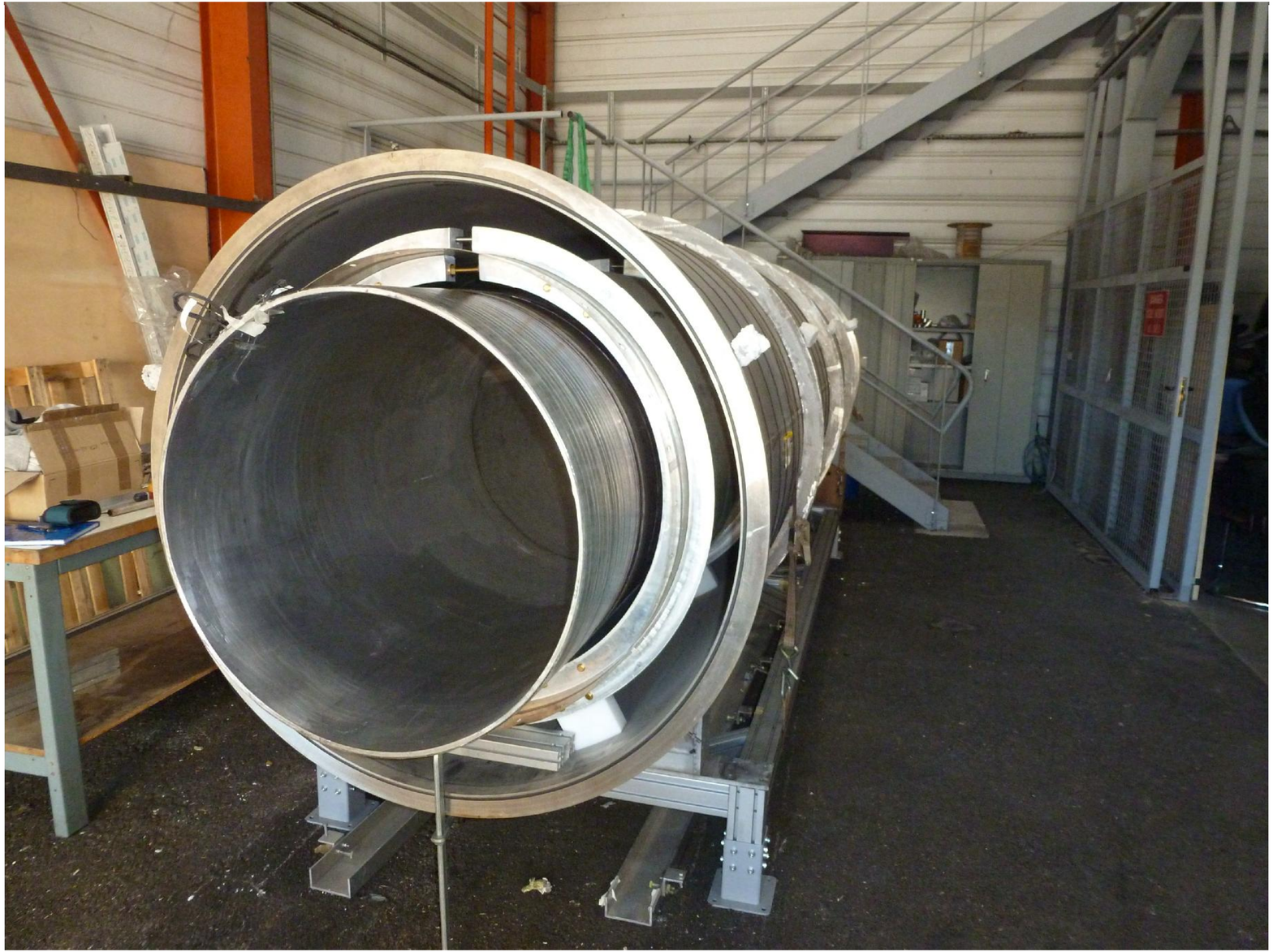}
		\includegraphics[width=0.470\textwidth]{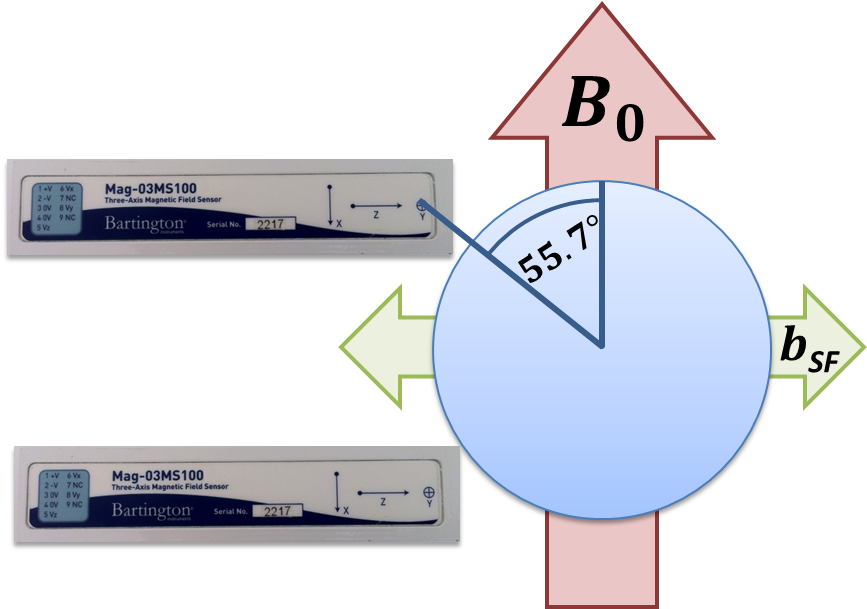}
	\caption{Left: Solenoid inserted into the mumetal magnetic shield. Right: Scheme of the spinflip and measurement apparatus. The spherical cell is positioned in contact with the two magnetometers inside the spinflip coil which generates an oscillating magnetic field $b_{SF}$ transversely to the holding magnetic field $B_0$.}
	\label{fig:vue_ensemble}
\end{figure}

In order to suppress the magnetic inhomogeneities responsible for the depolarization channel $\Gamma _m$, the holding field should be as homogeneous as possible.
The magnetic setup is composed of a $5~\rm{m}$ long and $80~\rm{cm}$ diameter solenoid which provides a very homogeneous magnetic field.
In order to shield the center of the solenoid where the cell will be set, a $\mu$-metal magnetic shield of $4.5~\rm{m}$ long and $96~\rm{cm}$ diameter was refurbished from the "n-nbar" experiment \cite{Bitter1991} which measured the neutron-antineutron oscillation at the ILL.
Fig. \ref{fig:vue_ensemble} shows the solenoid inserted into the $\mu$-metal magnetic shield.

We mapped the inner magnetic field using a three-axis fluxgate magnetometer, in order to extract the averaged squared transverse gradients $\langle (\vec{\nabla} b_{\perp})^2\rangle$ for different $B_0$ settings. Typically, $\sqrt{\langle (\vec{\nabla} b_{\perp})^2\rangle} $ is about $2.4~\rm{nT/cm}$ to $4~\rm{nT/cm}$ for a magnetic field between $2~\rm{\mu T}$ and $80~\rm{\mu T}$. 
The magnetic relaxation time $T_m = \frac{1}{\Gamma _m}$ is then expected to be longer than 90 h. 
This long time will directly affect the duration of the experiment and so the precision of the relaxation rate measurement.

\subsection{Measurement of the relaxation rates}

The spin relaxation of a gas can be determined by periodically measuring the polarization of the gas.
We used a direct polarimetry method, first presented by Cohen-Tanoudji \cite{Cohen-Tannoudji1969}, which consists in measuring the magnetic field generated by the gas itself and so proportional to its polarization.
Close to the cell, this magnetic field can be of the order of tenths of nT at pressures of several bars.
The sensors are set at a place where the dipolar field induced by the polarized gas is transverse relative to the $B_0$ field, as represented on Fig. \ref{fig:vue_ensemble}.
%\begin{figure}
	%\centering
		%\includegraphics[width=0.470\textwidth]{images/Image4.png}
	%\caption{Scheme of the spinflip and measurement apparatus. The spherical cell is positioned in contact with the two magnetometers inside the spinflip coil which generates an oscillating magnetic field $b_{SF}$ transversely to the holding magnetic field $B_0$.}
	%\label{fig:Image4}
%\end{figure}
The configuration with two fluxgates allows to compensate for the random fluctuations of the environmental transverse fields by taking the difference between the two magnetometer readings, as the dipolar fields created by the polarized $^3$He gas at positions 1 and 2 are opposite. 

\begin{figure}[htb]
	\centering
		\includegraphics[width=0.60\textwidth]{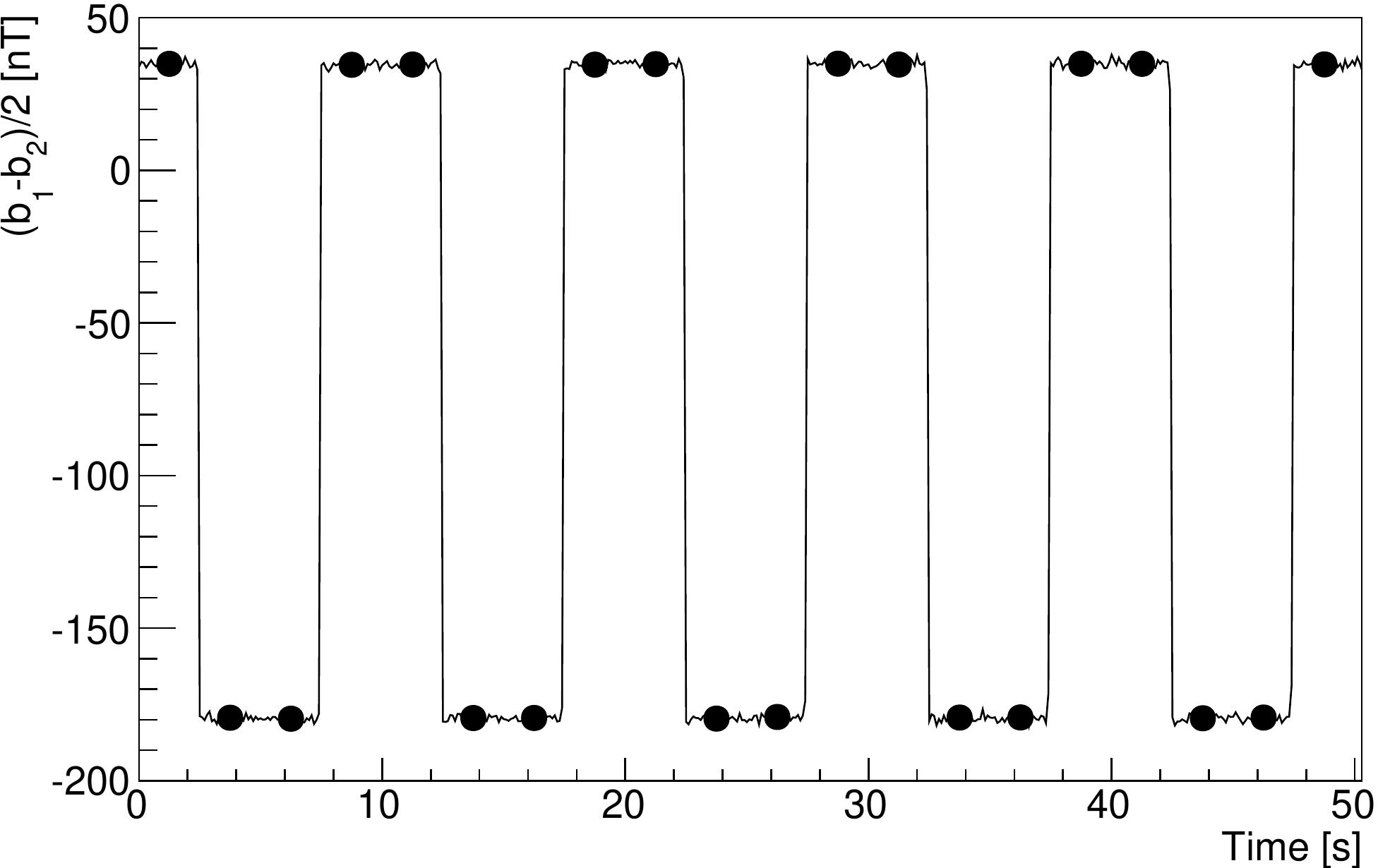}
	\caption{Typical sequence of measurement of the transverse magnetic field generated by a cylindrical cell at 4~bars with two fluxgate magnetometers. 
	The upper (lower) points correspond to the spin up (down) state of the gas.}
	\label{fig:MesurePedagogique}
\end{figure}

Applying spinflips with a transverse oscillating magnetic field to reverse the polarization (Fig. \ref{fig:vue_ensemble}), one can remove magnetic offsets induced by long-term drift of the holding magnetic field or the misalignment of the magnetometers axis with $B_0$. 
Fig. \ref{fig:MesurePedagogique} shows typical sequences of "up-down-down-up" (\uddu) measurements of the magnetic field induced by a spherical cell at 4~bar with the two magnetometers.
The precision on the polarization measurement extracted from the two fluxgates with a \uddu~sequence is $160~\rm{pT}$.
In order to improve this precision, we repeat the polarisation measurement 8 times.
For a $1~\rm{bar}$ $100~\%$ polarization cell, we have a signal over noise ratio about 1000.

%%%%%%%%%%%%%%%%%%%%%%%
\section{Data analysis}

In the experiment, we measure the relaxation rate of spin-polarized $^3$He gas as a function of the strength of the applied holding magnetic field $B_0$. 
To do so, we used spherical cells filled with polarized $^3$He.
The details of the 4 Runs obtained up to now are given in Table \ref{tab:liste-Runs}. 
\begin{table}[t]
	\centering
		\begin{tabular}{c|c|c|c|c|c}
		%Run			& Cell        & Radius & Pressure & Coating  &  Initial        & $g_sg_p\lambda$  (95 \% C.L.)    \\
		        %&             & [cm]   & [bar]    &          &  polarization   &  [$\mathrm{m}^2$]  \\ \hline
		%%31      & Axion 01 & 6		 & 1 	      & Caesium  & $34~\%$							\\
		%32      & Axion01     & 6		   & 1 	     & Caesium  & $70~\%$					&	$11\times 10^{-28}$		\\
		%33      & Axion01     & 6		   & 4 	     & Caesium	& $70~\%$					&	$6.7\times 10^{-28}$		\\
		%34      & CCT12       & 4	     & 4 	     & Rubidium & $70~\%$					&	$21\times 10^{-28}$	  \\
		%35      & CCT12       & 4		   & 1 	     & Rubidium & $70~\%$					&	$11\times 10^{-28}$	  %\\
		%36      & Axion01     & 6		  & 2 	     & Rubidium & $70~\%$						  \\
		%37      & Axion01     & 6		  & 3 	     & Rubidium & $70~\%$					    \\	
		%38      & BufferAspec & 6	    & 1 	     & No coating & $70~\%$					  \\	
		%39      & Axion01     & 6		  & 0.3 	   & Rubidium & $70~\%$						  \\
		%41      & Axion01     & 6		  & 3		 	   & Rubidium & $70~\%$		
		Run			& Cell        & Radius & Pressure & Coating  &  Initial         \\
		        &             & [cm]   & [bar]    &          &  polarization   \\ \hline
		%31      & Axion 01 & 6		 & 1 	      & Caesium  & $34~\%$							\\
		32      & Axion01     & 6		   & 1 	     & Caesium  & $70~\%$					\\
		33      & Axion01     & 6		   & 4 	     & Caesium	& $70~\%$					\\
		34      & CCT12       & 4	     & 4 	     & Rubidium & $70~\%$					\\
		35      & CCT12       & 4		   & 1 	     & Rubidium & $70~\%$					 %\\
		%36      & Axion01     & 6		  & 2 	     & Rubidium & $70~\%$						  \\
		%37      & Axion01     & 6		  & 3 	     & Rubidium & $70~\%$					    \\	
		%38      & BufferAspec & 6	    & 1 	     & No coating & $70~\%$					  \\	
		%39      & Axion01     & 6		  & 0.3 	   & Rubidium & $70~\%$						  \\
		%41      & Axion01     & 6		  & 3		 	   & Rubidium & $70~\%$		
		\end{tabular}
		\caption{Main characteristics of the 4 Runs obtained with the apparatus.
	The value of the indicated initial polarization was measured on the Tyrex installation using an optical method \cite{Andersen2005}.}
	\label{tab:liste-Runs}
\end{table}
For each Run, we applied 8 different values of holding field: $\left\{ 3~\rm{\mu T},\, 5~\rm{\mu T},\, 8~\rm{\mu T},\, 10~\rm{\mu T},\, 30~\rm{\mu T},\, 50~\rm{\mu T},\, 80~\rm{\mu T},\, 90~\rm{\mu T}\right\}$.
For a given holding field value, the polarization measurements were done every 20~min.
When the polarization decreased by $10~\%$, a different holding field value is used.

\begin{figure}
	\centering
		\includegraphics[width=0.60\textwidth]{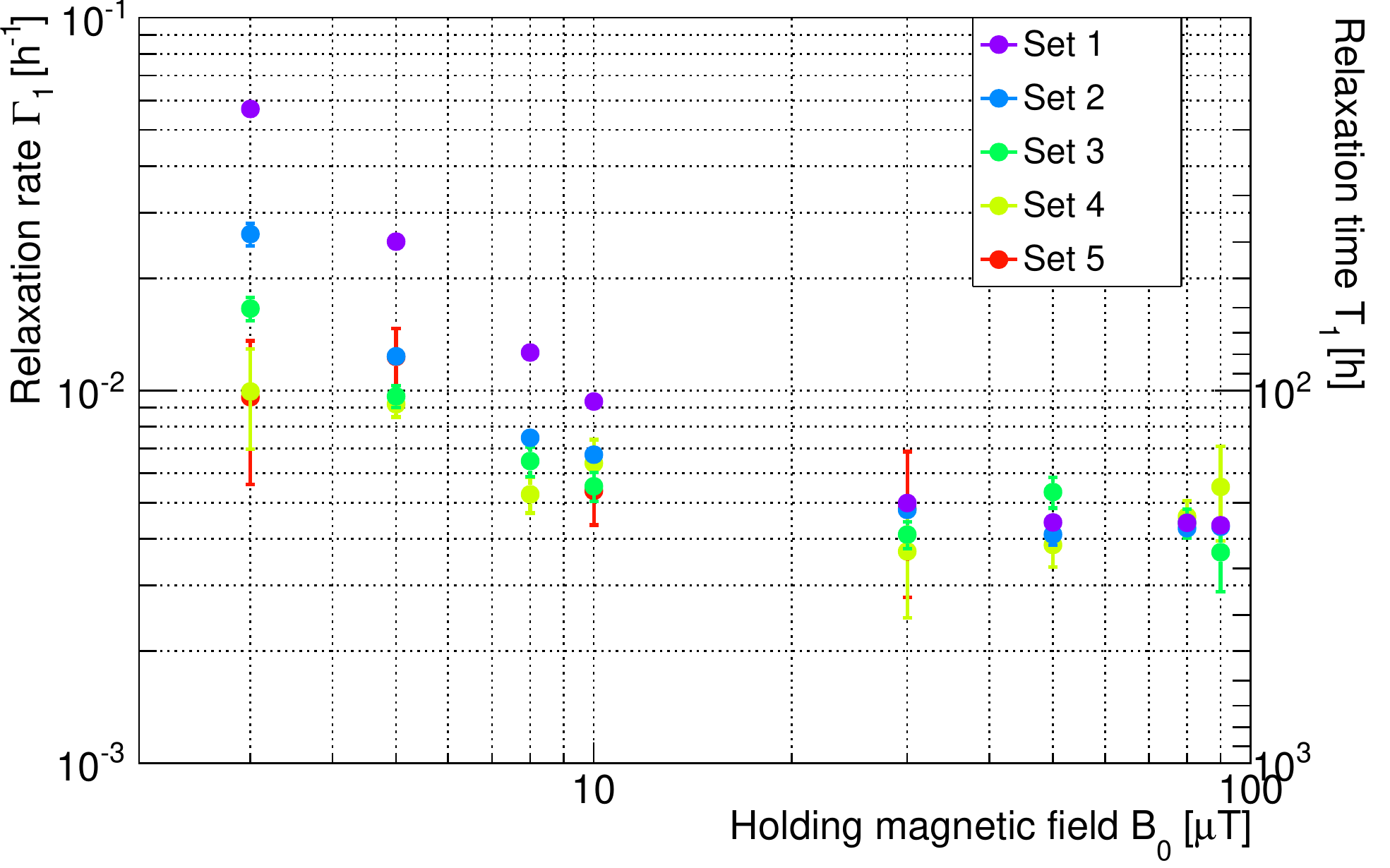}
	\caption{Measurement of the relaxation rates in the Run 32 as a function of the holding magnetic field.
	Each set corresponds to a series of holding field values.}
	\label{fig:Run32-GammavsB0-color}
\end{figure}

Using \eqref{eq:evolution-P}, we can extract the relaxation rate $\Gamma _1$ and the initial polarization for each magnetic configuration.
Fig. \ref{fig:Run32-GammavsB0-color} presents the measurement of the relaxation rate as a function of the holding field $B_0$ 
The relaxation rate at high magnetic field is about $200~\rm{h}$, which corresponds mainly to relaxation induced by the walls and atoms collisions.
At small magnetic field values ($3~\rm{\mu T}$), the relaxation is mainly due to gradients which are independent of the holding field value.
Since the relaxation rate decreases with the set number (i.e. as the gas depolarizes), the main depolarizing process at low magnetic field an high polarization comes from the magnetic gradients generated by the gas itself.

We can compare the behavior of the relaxation rate with the holding field and the holding field values with the expected standard rate \eqref{eq:Gamma1-explicite}.
If a new interaction  between two nucleons mediated by a pseudoscalar boson exists, a new depolarization channel \eqref{eq:Gamma1NF-formulefinale} will add to the standard one and could be detected as an depolarization excess.

However, this analysis method is not optimal since it does not take into account the correlation between the initial polarization value of each magnetic configuration. 
A new analysis method is currently under development in order to take into account these correlations.

\section{Conclusion}

Measuring $^3$He hyperpolarized gas relaxation as a function of the holding field is a very sensitive method to search for exotic short-range spin-dependent interactions.
A dedicated apparatus was built at the ILL in order to find a new spin-0 boson or to  improve the current constraint on its scalar-pseudoscalar couplings.
The setup is composed with a solenoid and a mumetal magnetic shield which provide a very homogeneous magnetic field.  
We hope to reach one order of magnitudeimprovement  in sensitivity compared with the previous experiment \cite{Petukhov2010}.
%allowed us to  set a very strong constraint on the scalar-pseudoscalar coupling constants of a spin-0 light boson to nucleons.

%%%%%%%%%%%%%%%%%%%%%%%%%%%%%%%%%%%%%%%%%%%%%%%%%%%%%%%%%%%%%%%%%%%%%%%%%%
%%%
%%%   use this format to include an .eps figure into your paper
%%%
%\begin{figure}[htb]
%\centering
%\includegraphics[height=1.5in]{}
%\caption{Plan of the magnet used in the mesmeric studies.}
%\label{fig:magnet}
%\end{figure}
%%%%%%%%%%%%%%%%%%%%%%%%%%%%%%%%%%%%%%%%%%%%%%%%%%%%%%%%%%%%%%%%%%%%%%%%%%%%

%%%%%%%%%%%%%%%%%%%%%%%%%%%%%%%%%%%%%%%%%%%%%%%%%%%%%%%%%%%%%%%%%%%%%%%%%%
%%%
%%%   use this format to include a LaTeX table  into your paper
%%%
%\begin{table}[t]
%\begin{center}
%\begin{tabular}{l|ccc}  
%Patient &  Initial level($\mu$g/cc) &  w. Magnet &  
%w. Magnet and Sound \\ \hline
 %Guglielmo B.  &   0.12     &     0.10      &     0.001  \\
 %Ferrando di N. &  0.15     &     0.11      &  $< 0.0005$ \\ \hline
%\end{tabular}
%\caption{Blood cyanide levels for the two patients.}
%\label{tab:blood}
%\end{center}
%\end{table}
%%%%%%%%%%%%%%%%%%%%%%%%%%%%%%%%%%%%%%%%%%%%%%%%%%%%%%%%%%%%%%%%%%%%%%%%%%%%

%\Acknowledgements

\end{document}